\documentclass[a4paper,11pt]{article}

\usepackage[table]{xcolor}
\usepackage[margin=1in]{geometry}
\usepackage{tabularx}
\usepackage{enumitem}
\setlist{nolistsep}
\definecolor{blue}{HTML}{4682B4}
\definecolor{blue}{HTML}{000080}

\usepackage{float}
\usepackage{pos}
\title{Reconstruction of highly inclined extensive air showers in GRAND}
\manuallySeparateAuthors
\author*[a, b]{Oscar Macias,}
\author[c,d]{Aur\'elien Benoit-L\'evy,}
\author[e]{Valentin Decoene,}
\author[c,f]{Ars\`ene Ferri\`ere,}
\author[f,g]{Marion Guelfand,}
\author[h]{Claire Gu\'epin,}
\author[g,i,d]{Kumiko Kotera,}
\author[a]{Zhisen Lai,}
\author[f,g]{Olivier Martineau-Huynh,}
\author[j]{Simon Prunet,}
\author[k,l]{and Matías Tueros,}
\author{   for the GRAND Collaboration}

\affiliation[a]{Department of Physics and Astronomy, San Francisco State University, San Francisco, CA 94132, USA}
\affiliation[b]{GRAPPA Institute, University of Amsterdam, Science Park 904, 1098 XH Amsterdam, Netherlands}
\affiliation[c]{Universit\'{e} Paris-Saclay, CEA, List, F-91120, Palaiseau, France}
\affiliation[d]{Department of Physics, Pennsylvania
State University, University Park, PA, USA}
\affiliation[e]{Subatech, IMT Atlantique, IN2P3-CNRS, Nantes Universit{\'e}, 4 rue Alfred Kastler - La Chantrerie, Nantes, BP 20722 44307 France}
\affiliation[f]{Sorbonne Universit\'{e}, CNRS, Laboratoire de Physique Nucléaire et des Hautes Energies (LPNHE),4 Pl. Jussieu, 75005 Paris, France}
\affiliation[g]{Sorbonne Universit\'{e} et CNRS, UMR 7095, Institut d'Astrophysique de Paris, 98 bis bd Arago, 75014 Paris, France}
\affiliation[h]{Laboratoire Univers et Particules de Montpellier (LUPM), Universit\'e de Montpellier, CNRS/IN2P3, CC72, Place Eug\`ene Bataillon, F-34095 Montpellier Cedex 5, France}
\affiliation[i]{Vrije Universiteit Brussel, Physics Department, Pleinlaan 2, 1050 Brussels, Belgium}
\affiliation[j]{Université Côte d’Azur, Observatoire de la Côte d’Azur, CNRS, Laboratoire Lagrange, \\ Bd de l’Observatoire, CS 34229, 06304 Nice cedex 4, France}
\affiliation[k]{IFLP - CCT La Plata - CONICET, Casilla de Correo 727 (1900) La Plata, Argentina}
\affiliation[l]{Depto. de F\'isica, Fac. de Cs. Ex., Universidad Nacional de La Plata, Casilla de Coreo 67 (1900) La Plata, Argentina}

\emailAdd{macias@sfsu.edu}

\abstract{The Giant Radio Array for Neutrino Detection (GRAND) aims to detect highly inclined extensive air showers (EAS) with down-going and up-going trajectories. Several working groups in the GRAND collaboration are developing methods to reconstruct the incoming direction, core position, primary energy, and composition of the showers. The reconstruction pipeline---currently under development in the France/IAP working group---relies on a model of spherical wavefront emission for arrival times, which is possible because the radio signals are generated far away from the antenna stations. The amplitude distribution of the signals at the antenna level is described by an Angular Distribution Function that considers various asymmetries in the data, including geomagnetic effects. In this contribution, we present preliminary results from testing our EAS reconstruction procedure using realistic mock observations.}

\FullConference{%
  ARENA 2024 -- 10$^{\rm th}$ International Workshop on Acoustic and Radio EeV Neutrino Detection Activities\\
  June 11-14, 2024\\
  KICP (University of Chicago), Chicago, USA
}

\begin{document}
\maketitle

\section{Introduction}

To enable neutrino astronomy, it is essential to determine their arrival direction with high accuracy. An angular resolution of $\sim 0.1^\circ$ would allow us to differentiate between nearby sources (e.g.,~\cite{Guepin:2022qpl}). This sensitivity is also crucial for directly distinguishing extensive air showers (EAS) caused by neutrinos from those caused by cosmic rays or gamma rays, as none of the latter are expected to originate from below the horizon~\cite{Decoene:2020yrq,Decoene:2021ncf}. Moreover, reconstructing the arrival direction is crucial in determining the characteristics of EAS, such as the species and energy of the primary particle.

The Giant Radio Array for Neutrino Detection (GRAND)~\cite{GRAND:2018iaj, kotera:2024grd} is a planned ultra-high-energy (UHE) observatory with extensive exposure at UHEs and exquisite angular sensitivity. The GRAND collaboration is developing and testing new reconstruction methods specialized for highly inclined EASs. Specifically, the collaboration is focused on three main tasks: (1) Creating a library of highly realistic air shower simulations, including signals with realistic noise---both instrumental and astrophysical in origin---that can be used to develop and test the new reconstruction pipelines. (2)  Employing empirical and physics-informed models for reconstruction, as well as (3) advancing various electric field reconstruction methods based on first principles, measured polarization, and denoising algorithms. See Table~\ref{Tab:overview} for an overview. In the following, we will highlight some preliminary results and advances on these topics.

\begin{table}[b!]
\centering
\begin{tabularx}{\textwidth}[t]{X}
\arrayrulecolor{blue}\hline
\textbf{\textcolor{blue}{Reconstruction of highly-inclined Air Showers (conventional + ML methods)}}  \\
\hline
\end{tabularx}
\begin{tabularx}{\textwidth}[t]{XX}
1. Realistic Data Simulation libraries & 
\begin{minipage}[t]{\linewidth}%
\begin{itemize}
\item[1.1] Use both ZHAireS and CoREAS.
\item[1.2] Include  Galactic noise + antenna response + RF chain + GPS
jitter
\end{itemize}
\vspace{0.2mm}
\end{minipage}\\
\arrayrulecolor{black}\hline

2. Reconstruction of Air Showers &
\begin{minipage}[t]{\linewidth}%
\begin{itemize}
\item[2.1] Plane Wavefront (PWF) using timing and
antenna positions~\cite{Decoene:2021ncf}
\item[2.2] Fitting empirical and Physics informed~\cite{gulzow:2024grd} models for the Angular Distribution Function (ADF)~\cite{Decoene:2020yrq}
\item[2.3] Empirical fitting of lateral distribution function
\item[2.4] Graph Neutral Networks for EAS studies
\end{itemize}
\vspace{0.4mm}
\end{minipage}\\

\hline

3. Electric field reconstruction and triggering algorithms &
\begin{minipage}[t]{\linewidth}%
\begin{itemize}
\item[3.1] Novel trigger algorithms~\cite{koler:2024grd, correa:2024grd} 
\item[3.2] E-field reconstruction with CNNs
\item[3.3] Direction reconstruction based on
polarization
\item[3.4] Denoising of E-field/ADC using ML methods
\end{itemize}
\vspace{0.4mm}
\end{minipage}\\
\arrayrulecolor{blue}\hline
\end{tabularx}
\caption{Overview of the three main tasks and their respective subtasks being developed within the GRAND collaboration to address the problem of reconstructing highly-inclined air showers using conventional and machine learning methods.}
\label{Tab:overview}
\end{table}
\begin{figure}[t!]
    \centering
    \tabular{ll}
    \includegraphics[width=0.5\linewidth]{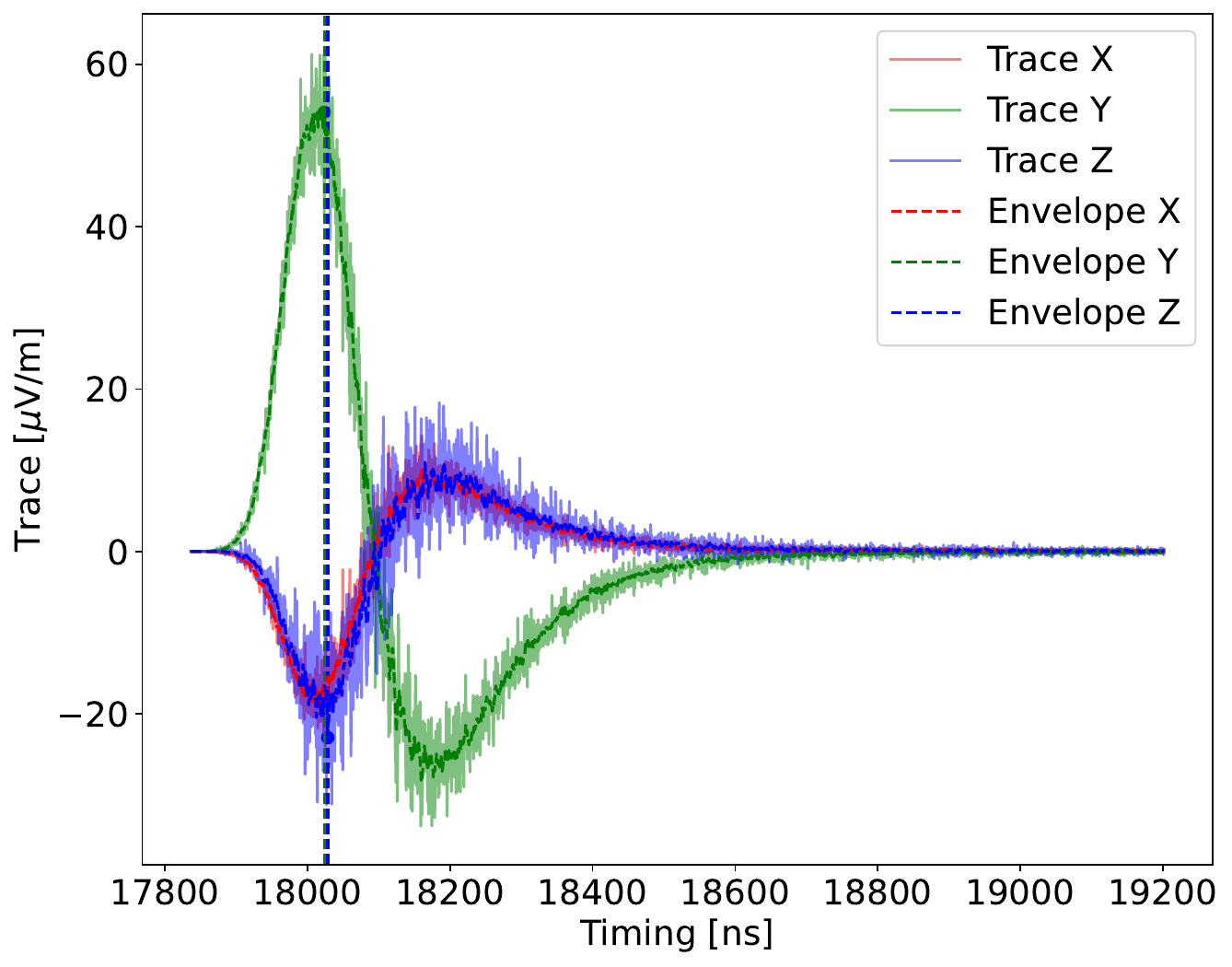} & 
    \includegraphics[width=0.5\linewidth]{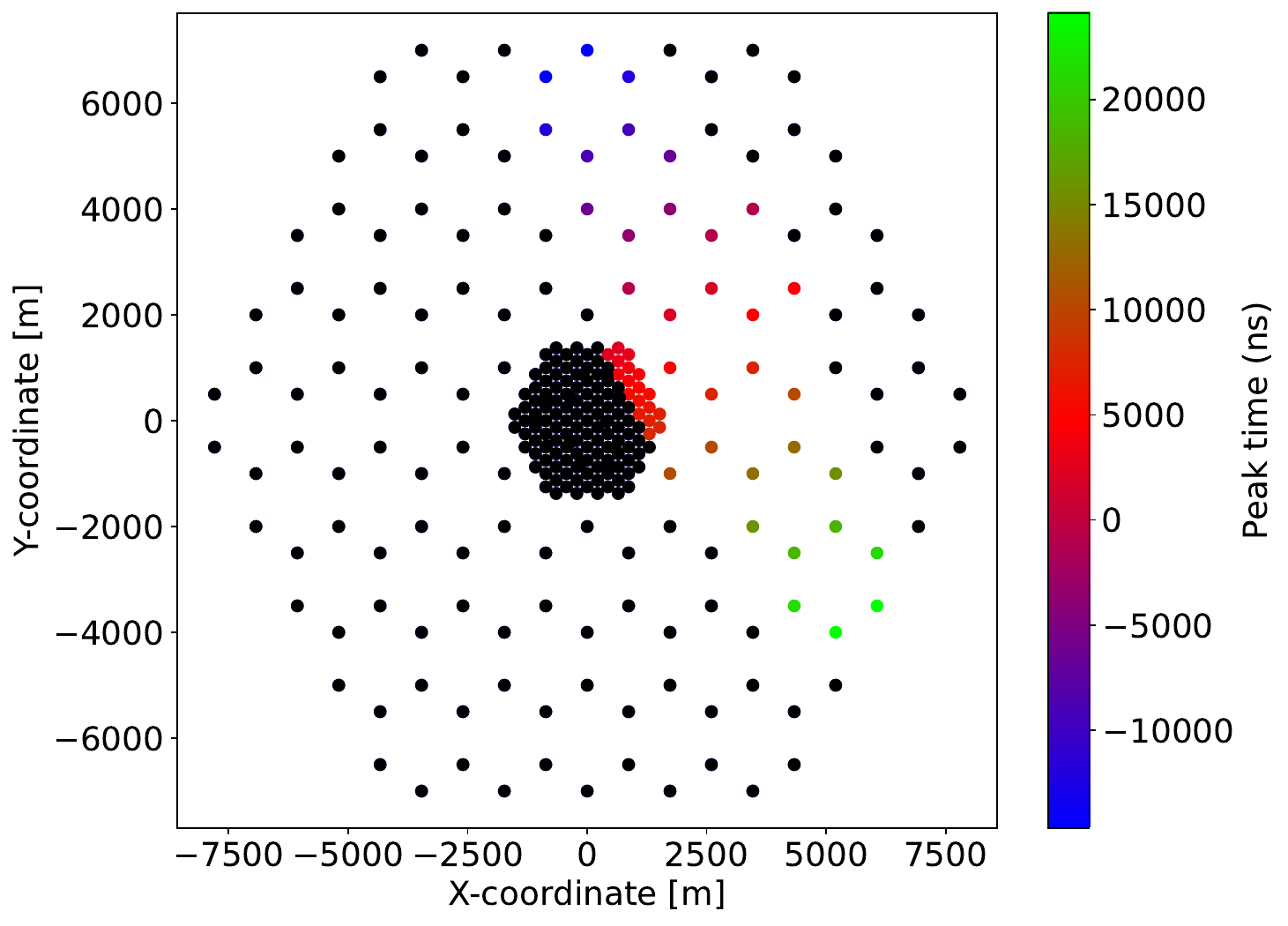}
    \endtabular
    \caption{Realistic simulation of an UHE proton with an energy of 2.81 EeV, and zenith and azimuth angles of $67.05^\circ$ and $22.51^\circ$, respectively. This simulation is part of the \textit{``Data Challenge 2''} (DC2) library of simulations used to validate the reconstruction algorithms being developed within the collaboration. \textbf{Left:} electric field amplitude for the three different polarizations along the $X$, $Y$, and $Z$ directions. The envelope displays the mean of the signal taken in time windows of 10 ns. \textbf{Right:} antenna positions (in the flat Earth approximation). Black markers represent untriggered antennas, while the blue-to-green markers indicate the trigger time in ns. The average distance between the antennas is approximately 1 km, except for the infill at the center, where the distance is less than 250 m.   }
    \label{fig:sims}
\end{figure}

\section{Realistic Data Simulation Libraries}
\label{Sec:Simulations}

The ``Data Challenge 2'' (DC2) features a grid of 289 radio antennas in China's Xiadoushan region. These antennas form a hexagonal pattern with a 1000 m spacing and a denser infill area where the spacing is 250 m. The simulations were created using both the \texttt{ZHAireS}~\cite{zhaires:2012} and \texttt{CoREAS}~\cite{Huege:2013vt} software packages. Various actions were taken to ensure that the output from both packages had a compatible format. The simulations can be readily accessed and processed using the publicly available \texttt{GRANDLib} package~\cite{GRAND:2024atu}. The goal of this library is to be expansive and adaptable, meeting the needs of a diverse range of studies and emphasizing the development and testing of machine learning applications. In machine learning, it is essential to consistently cover a wide range of the model parameter space. To accomplish this, we use continuous parameter distributions, including a distinctive zenith angle distribution that smoothly encompasses all different footprint sizes. The full details of the simulations will be available in an upcoming publication.


\begin{figure}[t!]
    \centering
    \includegraphics[width=1\linewidth]{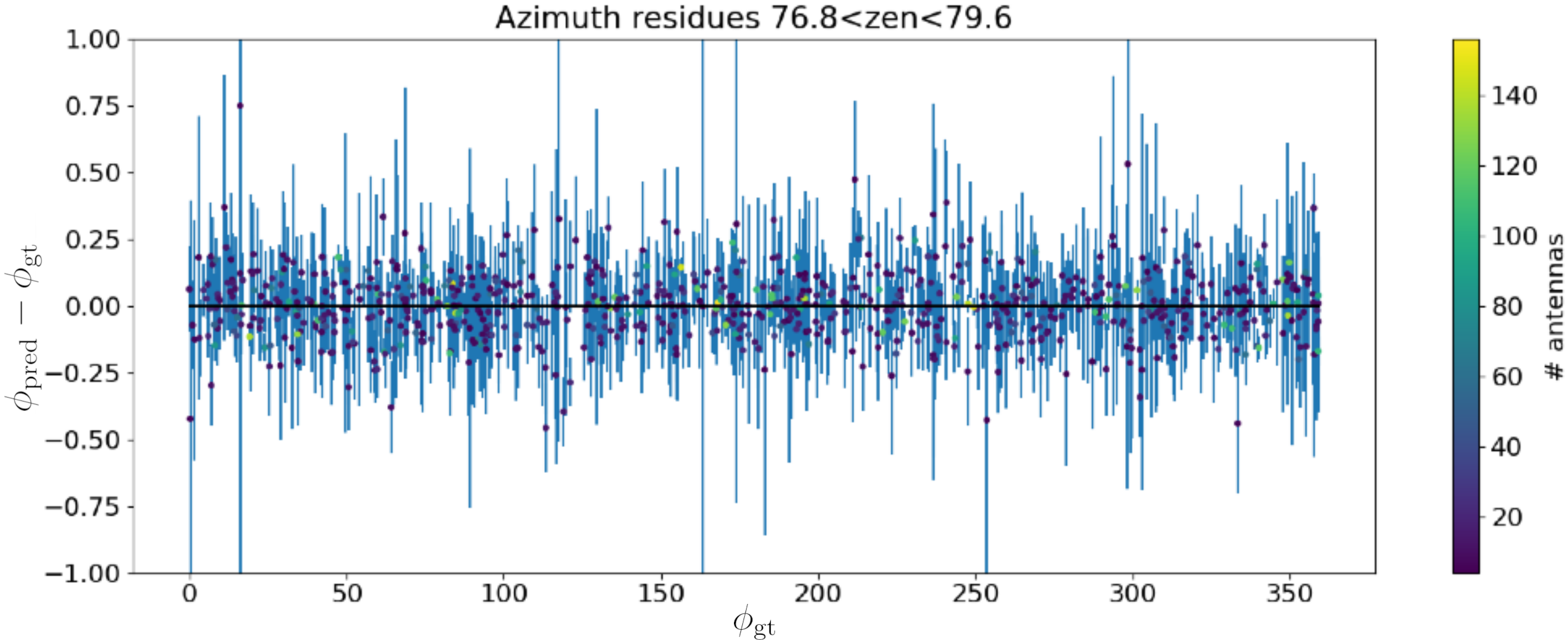}
    \includegraphics[width=1\linewidth]{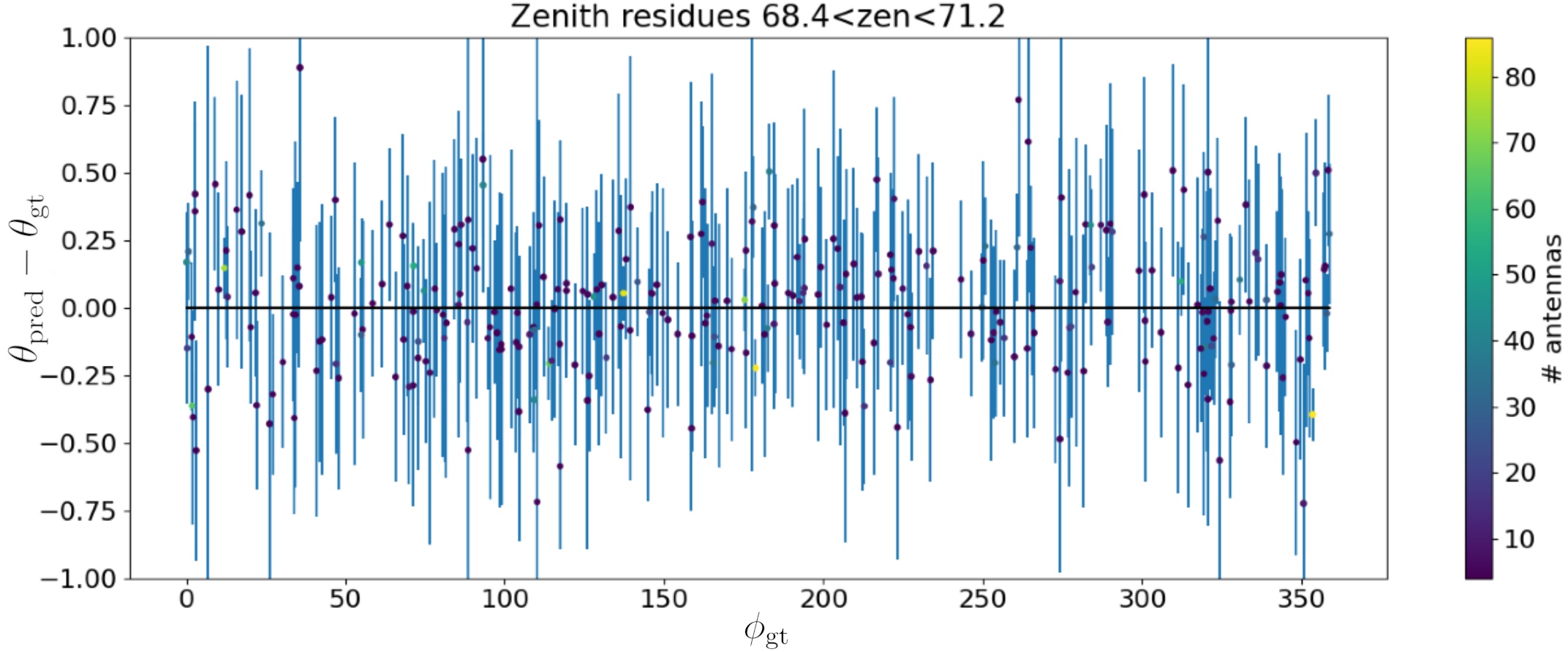}
    \caption{Reconstructed direction of UHECR events using a plane wavefront (PWF) model~\cite{Ferriere:2024bt}. The horizontal axis represents the ground truth (gt) azimuth, whereas the vertical axis shows the reconstructed residuals for the azimuth and zenith, respectively. Each point represents a separate simulation, and the color bar indicates the number of triggered antennas during the event. See Sec.~\ref{Sec:Simulations} for a description of the simulations.}
    \label{fig:pw}
\end{figure}

\begin{figure}[t!]
    \centering
    \includegraphics[width=0.7\linewidth]{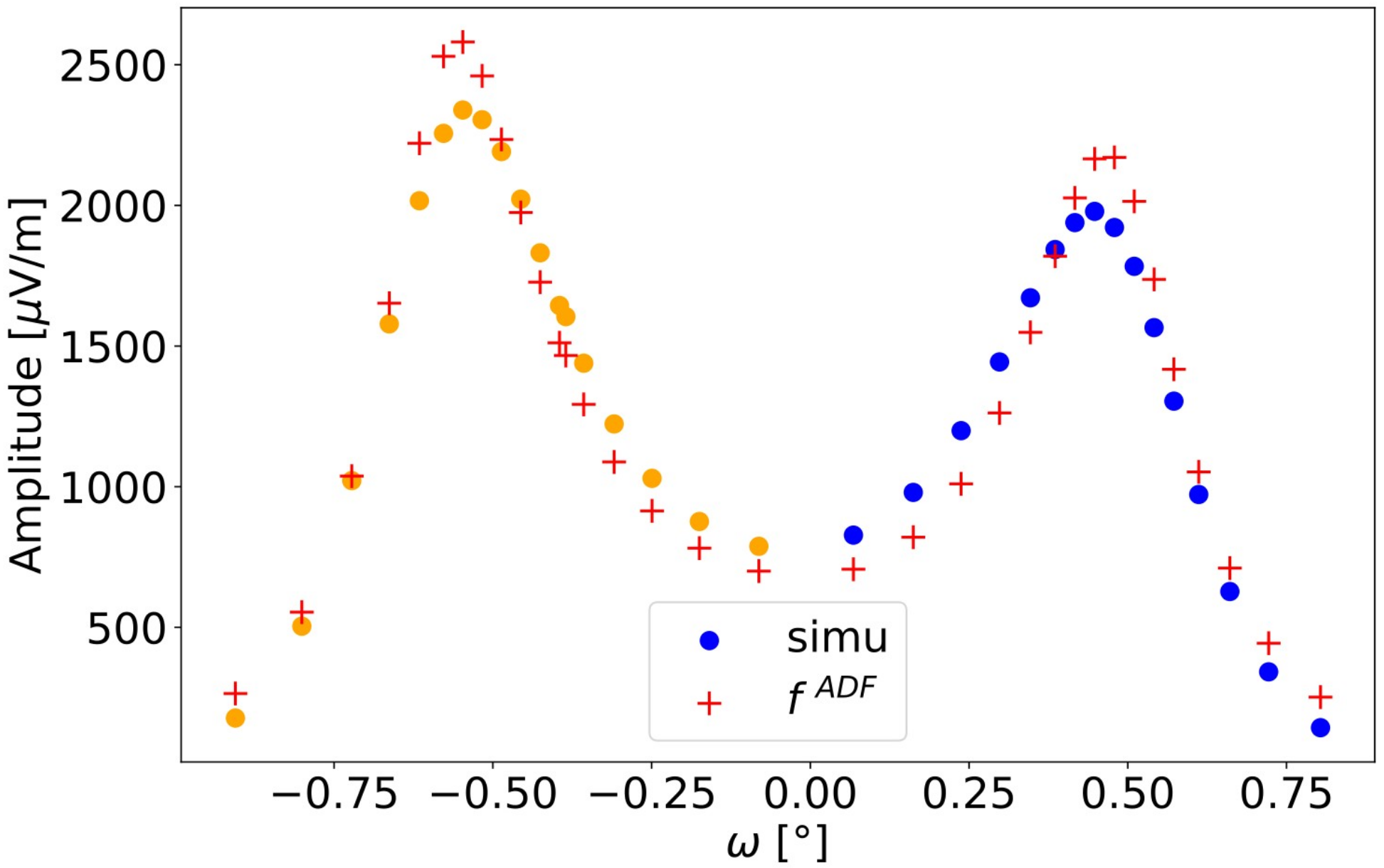}
    \caption{Comparison of the simulated versus reconstructed angular distribution function (ADF) for the electric field amplitude in a GP300~\cite{Decoene:2019sgx} simulation. The horizontal axis shows the shower angle $\omega$ measured from the shower axis, and the vertical axis is the E-field amplitude in the shower plane. }
    \label{fig:ADF}
\end{figure}

\section{Reconstruction of Air Showers in GRAND}
\label{Sec:Reconstruction}

Below, we highlight the preliminary results of two selected reconstruction procedures based on standard and machine learning methodologies being developed within the collaboration.

\subsection{Empirical reconstruction method:}
\label{subsec:empiricalreconstructionmethod}
This method, first introduced in Refs.~\cite{Decoene:2020yrq,Decoene:2021ncf}, combines arrival time and signal amplitude to reconstruct the emission point, arrival direction, and amplitude in a step-by-step process. Detected arrival times are initially used to estimate the emission point assuming a plane wavefront (PWF) model, followed by a refinement using a spheroidal one. Then, using this reconstructed emission point and the recorded amplitude pattern on the ground, the position of the shower axis is determined, thus establishing the direction of the EAS. This method considers asymmetry features and signal attenuation during propagation~\cite{Decoene:2021ncf}. Additionally, this approach enables the determination of the EAS's energy and composition by examining the event's geometry.

Figure~\ref{fig:pw} illustrates a comparison between the true and reconstructed directions of simulated UHECR events using only the PWF model (first step) of the empirical method. We considered only simplified (Gaussian) errors for this test~\footnote{Further details about our PWF implementation can be found in Ref.~\cite{Ferriere:2024bt}}. Similarly, Fig.~\ref{fig:ADF} compares the angular distribution function (ADF) for a GRAND-Proto-300 (GP300)~\cite{chiche:2024grd,Decoene:2019sgx} simulated event versus the reconstructed one. Consistent with previous findings presented in Ref.~\cite{Decoene:2020yrq,Decoene:2021ncf}, the pipeline demonstrates great potential, particularly for the most highly inclined events.

\begin{figure}[t!]
    \centering
    \tabular{ll}
    \includegraphics[width=0.5\linewidth]{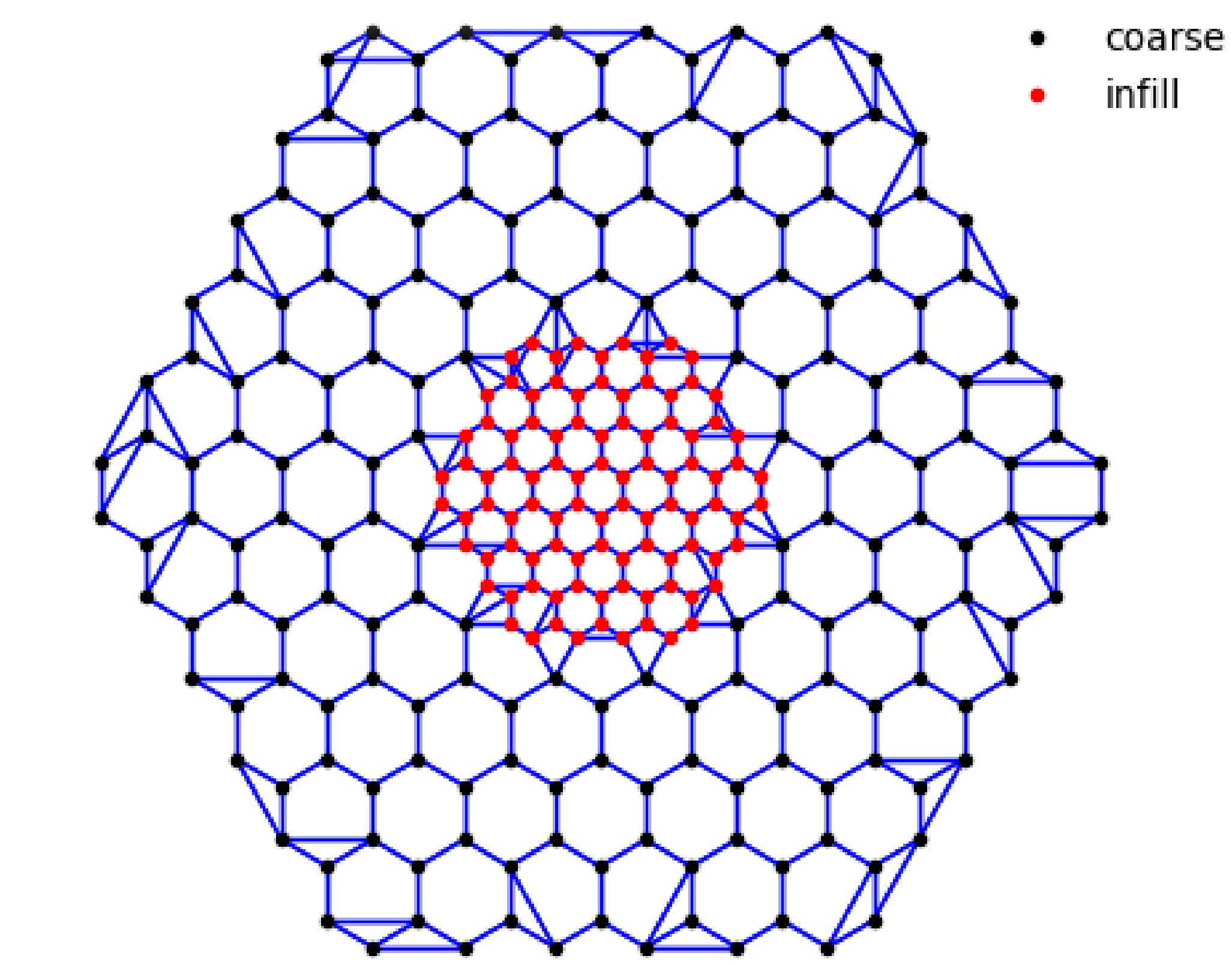} &
    \includegraphics[width=0.5\linewidth]{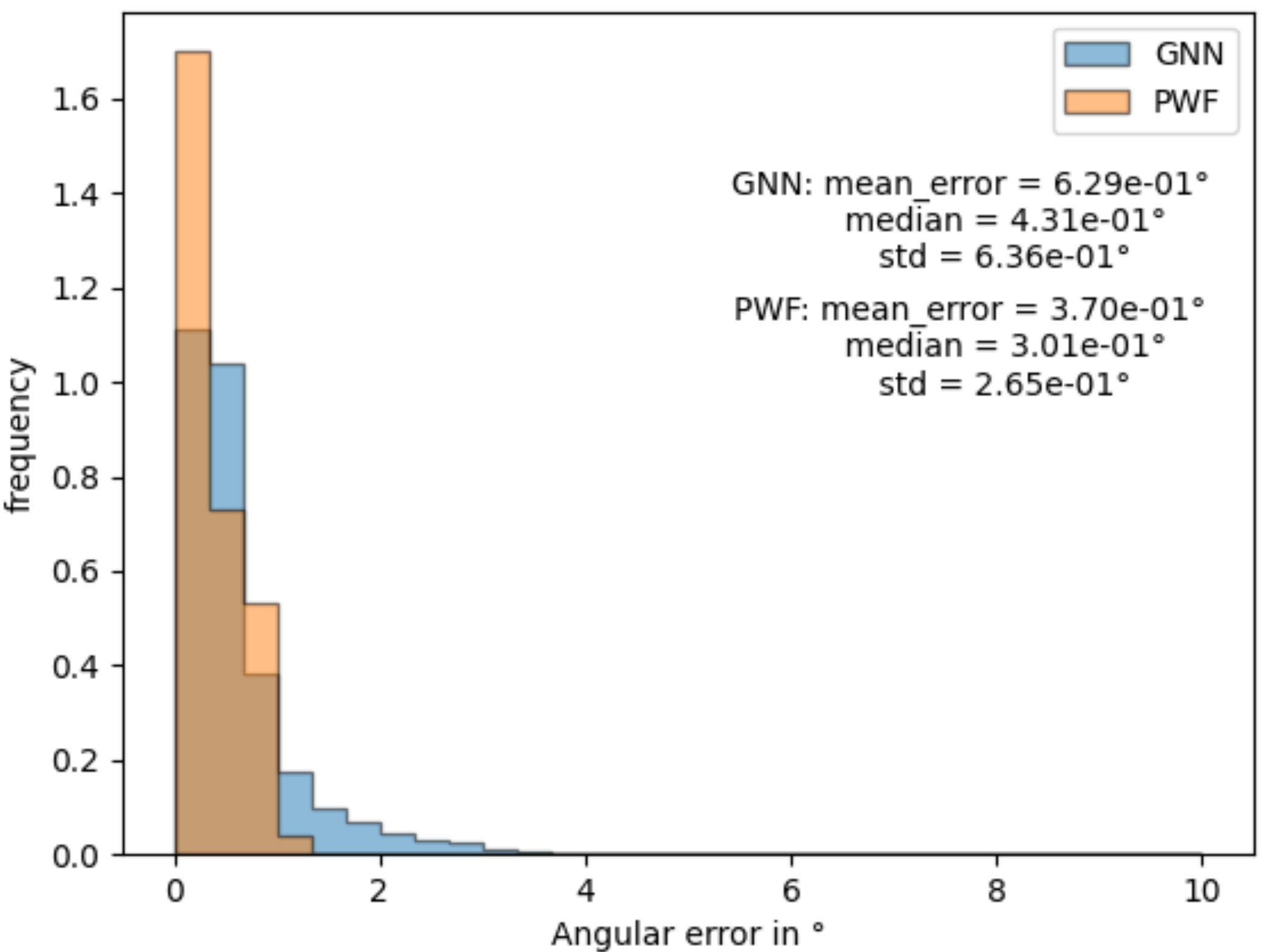}
    \endtabular
    \caption{Preliminary results of a direction reconstruction method based on a Graph Neural Network (GNN). \textbf{Left:} antenna array configuration based on the ``Data challenge 2'' simulations (see text). The black dots show the antenna positions (nodes), and the blue lines represent the connections between the nodes that make up the graph. \textbf{Right:} comparison between the direction resolution obtained with the GNN algorithm versus the Plane Waveform (PWF) model shown in Fig.~\ref{fig:pw}. Further improvements would allow matching the angular resolution of the more established PWF approach. }
    \label{fig:GNNs}
\end{figure}

\section{Reconstruction procedure based on Machine Learning algorithms}
\label{sec:machinelearning}

Extensive simulation studies (see Sec.~\ref{Sec:Simulations}) demonstrate that UHECR events create distinct antenna illumination patterns, encompassing electric field amplitudes and trigger times. This geometrical information is input into a Graph Neural Network (GNN) algorithm to represent GRAND events as point cloud graphs and utilize GNN for reconstruction. The model architecture, training configuration, and loss functions will be the subject of an upcoming publication. For now, we highlight preliminary results in Fig.~\ref{fig:GNNs} showing promising capabilities close to the more mature ADF algorithm.

\begin{figure}[t!]
    \centering
    \includegraphics[width=1.0\linewidth]{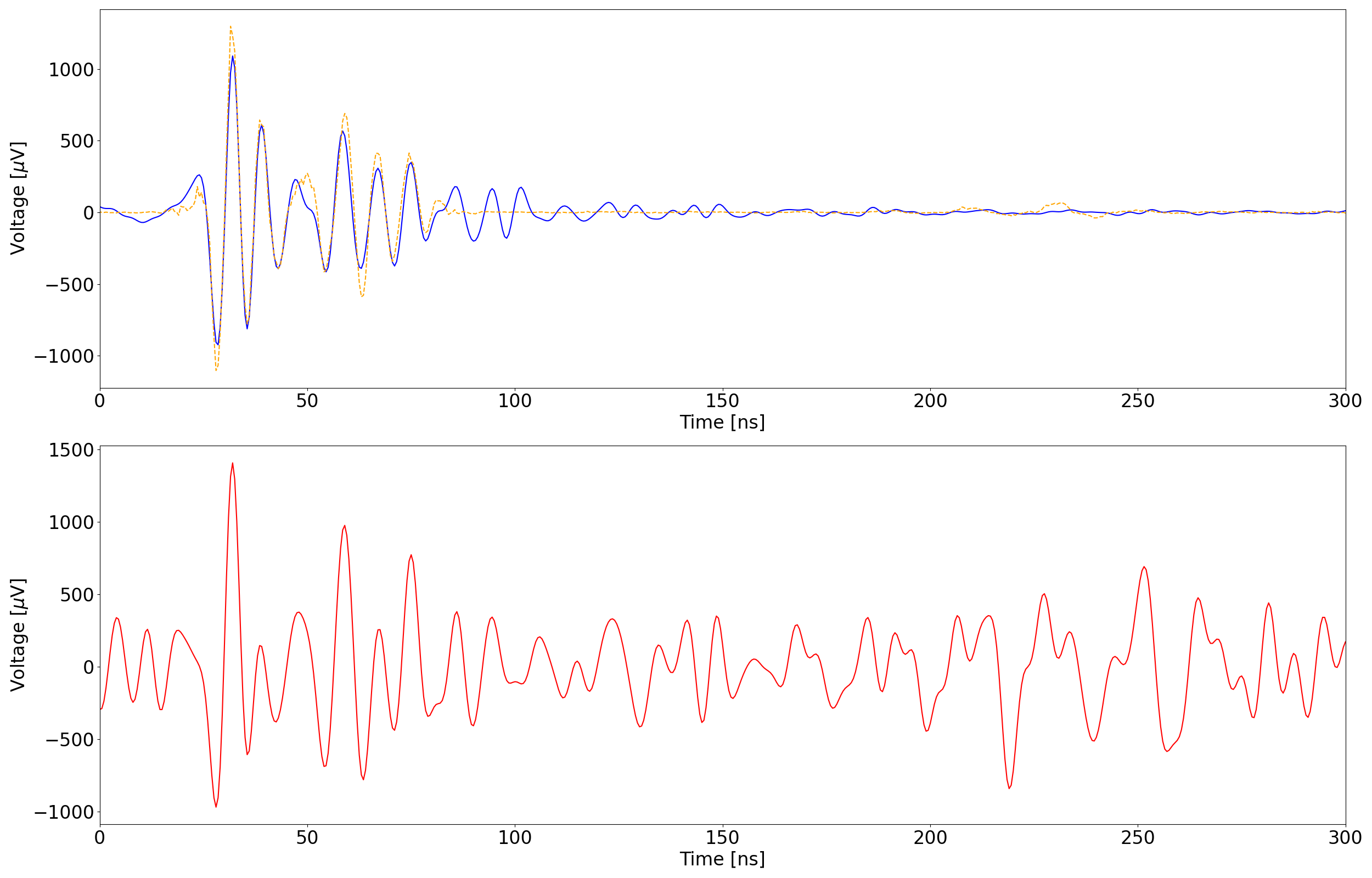}
    \caption{Preliminary results of a signal denoiser algorithm. \textbf{Top panel:} The blue line represents the simulated pure voltage signal at the antenna level. In contrast, the yellow line shows the result of passing the noisy signal in the bottom panel through a machine learning-based autoencoder algorithm.  \textbf{Bottom panel:} Realistic noisy voltage signal extracted from the DC2 library of simulations (see Sec.~\ref{Sec:Simulations}) with a signal-to-noise ratio (SNR) of 4.4. The autoencoder, trained with thousands of signals without noise, can remove most of the noise and accurately reconstruct the peak time and amplitude.}
    \label{fig:denoiser}
\end{figure}

\section{Denoising of voltage signals using machine learning methods}
\label{sec:denoiser}

Autoencoders are potent machine learning algorithms recently employed for denoising data in the field of radio astronomy. It encodes data into a small latent space and then decodes it back again, eliminating unwanted features such as noise. The GRAND collaboration is developing a denoising pipeline using an autoencoder. It is trained using thousands of DC2 simulations containing voltage traces at the antennas at the antenna level. Figure~\ref{fig:denoiser} shows an example of a voltage signal with an ${\rm SRN}=4.4$. It is difficult to identify where the signal is visually, but the autoencoder can recover most of the signal. This enables it to recover the trigger time and the peak amplitude accurately. We have obtained preliminary promising results for SNRs greater than 3, but further validation work is required to confirm the method's effectiveness fully.

\section{Conclusions}

The GRAND collaboration is advancing various reconstruction methods, which show promising results in preliminary testing with realistic mock observations. A combination of empirical methods where the physics is clearly understood and modern methods based on machine learning will allow us to gain efficiency and perform the necessary cross-checks for the reliable identification of the properties of UHE particles.

\bibliographystyle{aasjournal}
\bibliography{references}

\end{document}